# Realistic head modeling of electromagnetic brain activity: An integrated Brainstorm pipeline from MRI data to the FEM solution

*Takfarinas Medani[1], Juan Garcia-Prieto[2], Francois Tadel[1], Sophie Schrader[3,4], Anand Joshi[1], Christian Engwer[4], Carsten H. Wolters[3], John C. Mosher[2], and Richard M. Leahy[1]*

[1] Signal & Image Processing Institute, University of Southern California, Los Angeles, CA 90089, USA

[2] Department of Neurology, McGovern Medical School, University of Texas Health Science Center at Houston, Houston, TX, USA

[3] Institute for Biomagnetism and Biosignalanalysis, Westfälische Wilhelms-Universität Münster, Münster, Germany

[4] Department of Applied Mathematics, University of Münster, Germany

*ABSTRACT*

*Human brain activity generates scalp potentials (electroencephalography – EEG), intracranial potentials (iEEG), and external magnetic fields (magnetoencephalography – MEG), all capable of being recorded, often simultaneously, for use in research and clinical applications. The so-called* forward problem *is the modeling of these fields at their sensors for a given putative neural source configuration. While early generations modeled the head as a simple set of isotropic spheres, today's ubiquitous magnetic resonance imaging (MRI) data allows detailed descriptions of head compartments with assigned isotropic and anisotropic conductivities. In this paper, we present a complete pipeline, integrated into the Brainstorm software, that allows users to generate an individual and accurate head model from the MRI and then calculate the electromagnetic forward solution using the finite element method (FEM). The head model generation is performed by the integration of the latest tools for MRI segmentation and FEM mesh generation. The final head model is divided into five main compartments: white matter, grey matter, CSF, skull, and scalp. For the isotropic compartments, widely-used default conductivity values are assigned. For the brain tissues, we use the process of the effective medium approach (EMA) to estimate anisotropic conductivity tensors from diffusion weighted imaging (DWI) data. The FEM electromagnetic calculations are performed by the DUNEuro library, integrated into Brainstorm and accessible with a user-friendly graphical interface. This integrated pipeline, with full tutorials and example data sets freely available on the Brainstorm website, gives the neuroscience community easy access to advanced tools for electromagnetic modeling using FEM.*

***Keywords***: *realistic head model, Brainstorm, DUNEuro, brain activity, forward problem, EEG/MEG, FEM*

## *METHOD*

The solution of the electromagnetic *forward model* (Mosher et al., 1999) can be divided into two primary steps. The first is the generation of a realistic and individualized head model. The second is the calculation of the signal at the sensors for given brain activity within this head model. The two steps yield a "lead field matrix" that can be integrated into Brainstorm[a] (or other programs) for investigations of the *inverse problem* using source localization or cortical current density mapping to infer neural activity from observed sensor data.

### *Head model generation*

In the first primary step, we generate the head model from magnetic resonance imaging (MRI) data. T1 weighted (and, if available, T2) images are segmented into five main compartments: white matter, grey matter, cerebrospinal fluid (CSF), skull, and scalp. Segmentation is performed using the SPM toolbox (Friston, 2003), as called from within the Brainstorm pipeline. Depending on options selected by the user, either a tetrahedral or a hexahedral mesh is generated, using methods such as constrained Delaunay triangulation. Other tools are also employed by the pipeline, such as headreco, distributed with the SimNibs software (Saturnino et al., 2019), the brain2mesh toolbox (Tran et al., 2020), and the head reconstruction process from the Roast toolbox (Huang et al., 2019), all integrated as calls within the Brainstorm graphical interface and batch editor.

---

[a] https://neuroimage.usc.edu/brainstorm

*Brain tissue conductivity and anisotropy*

Each of the head model compartments requires realistic conductivity values. For the isotropic compartments (grey matter, CSF, skull, and scalp) the widely-used conductivity values from the literature are assigned by default for each compartment. For anisotropic tissues, especially white matter, the user can generate individual conductivity tensors from diffusion-weighted imaging (DWI) data. For this purpose, Brainstorm calls the BrainSuite Diffusion Pipeline[b] (BDP) (Bhushan et al., 2015; Shattuck and Leahy, 2002), which is used to estimate the diffusion tensors; the *effective medium approach* (EMA) is then used to compute the conductivity tensors (Haueisen et al., 2002; Tuch et al., 2001; Wolters et al., 2001) for each of the elements belonging to white matter.

*Finite element modeling (FEM)*

For the second primary step, the FEM computation, we use the DUNEuro library[c] (Nüßing et al., 2019). DUNEuro offers modern FEM methods such as Continuous and Discontinuous Galerkin FEM (Engwer et al., 2017; Nüßing et al., 2016; Piastra et al., 2018) as well as unfitted FEM (Nüßing et al., 2016; Piastra et al., 2018) with a variety of FEM source models (Medani et al., 2015; Miinalainen et al., 2019; Wolters et al., 2008). The FEM process in the Brainstorm pipeline follows a guided process similar to the other approaches already available within Brainstorm (Tadel et al., 2011), providing the user with default values from the literature for most of the FEM parameters. For experienced or curious users, an advanced panel is made available to tune the different FEM parameters, for example for more individual conductivity modeling (Antonakakis et al., 2019). At this time, both MEG and EEG are fully tested and validated in Brainstorm. Full documentation and dataset can be found on the Brainstorm website.

## RESULTS

We tested this pipeline on published data collected from a healthy adult subject: T1w, T2w, DWI, EEG, and MEG (Piastra et al., 2020). For the head model generation, the headreco process is used to construct the tetrahedral FEM mesh. Figure 1(a, c, and d) shows the obtained head model with the five compartments. For white matter anisotropy, the DWI data is processed with BDP, and the EMA is applied. Figure 1(b) shows the distribution of conductivity tensors as an ellipsoid on each FEM element. The orientation of the first tensor eigenvector is color-coded as follows: red for right-left, green for anterior-posterior, and blue for superior-inferior. Figure 1(c) shows the three modalities, the MRI, the FEM mesh, and the conductivity tensors as ellipsoids, overlaid on the same image.

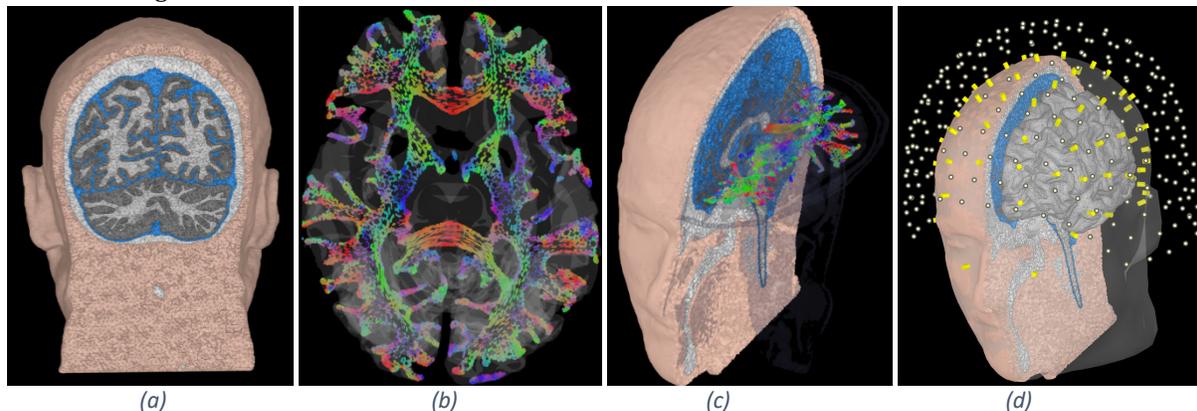

*(a)      (b)      (c)      (d)*

*Figure 1 The realistic FEM head model. (a) The tetrahedral FEM mesh with five compartments: white matter, grey matter, CSF, skull, and scalp. (b) The FEM tensors on the white matter as a color-coded ellipsoid computed from the DWI with BDP and the EMA. (c) The MRI, the FEM mesh, and the FEM tensors are overlaid on the same figure. (d) The FEM head model, the cortex (source space) and the location of the EEG (yellow) and MEG (white) sensors*

---

[b] http://brainsuite.org/processing/diffusion/
[c] http://duneuro.org/

The source space (location of the dipoles) used for the forward computation is obtained from the nodes of the cortical surface and corrected following Venant's condition (Medani et al., 2015; Wolters et al., 2008) using Brainstorm. For the final step of the FEM computation, the EEG and MEG sensors are aligned to the anatomy in Brainstorm, then the lead fields for these sensors computed throughout the locations of the source space. The result is a "lead field matrix", where the number of rows equals the number of sensor channels, and the number of columns equals the number of elemental sources, such that we have a mapping of every cortical dipole to every sensor.

Figure 1(d) shows the full head model with the EEG/MEG sensor location and the source space, for 74 EEG electrodes and 275 MEG sensors, and the source space comprising 15000 cortical dipoles. Figure 2 shows the distribution of the lead field for an EEG electrode pair (colored red and green), and a single MEG sensor in red.

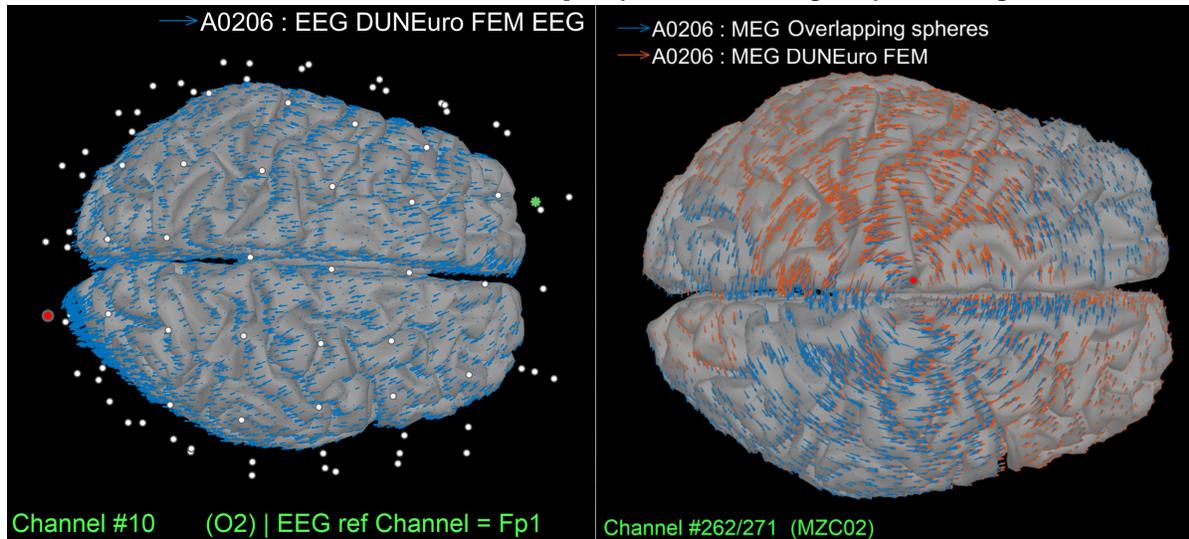

*Figure 2 Visualization of lead field vectors from Brainstorm: (left) the lead field vectors for a selected pair of channels that form the lead (FP1 in green, and O1 in red) for EEG; (right) the MEG lead field obtained with FEM (red arrows) and the overlapping spheres method available within Brainstorm (blue arrows).*

For comparison to a different head model, we computed the MEG solution for the same subject using the overlapping spheres (OS) method (Huang et al., 1999) available within Brainstorm (Tadel et al., 2011). Results are shown in Figure 2 (right). Both methods, FEM and OS, show good concordance. More detailed comparisons are under investigation and will be published in the near future.

## CONCLUSION

This paper describes the new full FEM pipeline integrated into Brainstorm, a software environment for neuroimaging data analysis. The software, documentation, and example datasets are freely available at https://neuroimage.usc.edu/brainstorm. The pipeline handles all the steps, from the processing of MRI data for individual and realistic head model construction to accurate FEM computations and advanced visualization. This integrated pipeline, with full tutorials and example data sets available on the website, gives the neuroscience community easy access to advanced tools for electromagnetic modeling with the FEM, implementing tools in the DUNEuro library across multiple platforms.

### Acknowledgment
The research reported in this publication was supported by the National Institute of Biomedical Imaging and Bioengineering (NIBIB) under award numbers R01EB026299 and U01EB023820. Its contents are solely the responsibility of the authors and do not necessarily represent the official views of the NIBIB.